\documentclass[reprint, aps,floatfix,showpacs,superscriptaddress]{revtex4-1}

\usepackage{graphicx,epsf,epsfig}
\usepackage{subfigure}
\usepackage{dcolumn}
\usepackage{bm}
\usepackage{ulem}
\usepackage{color}
\usepackage{amsmath}
\usepackage{amssymb}
\usepackage{mathptmx}      
\usepackage{latexsym}
\usepackage{verbatim}
\usepackage[english]{babel}
\usepackage{epstopdf}

\begin{document}

\title{Manipulating chiral transmission by gate geometry: switching in graphene \\with transmission gaps}
\author{Redwan N. Sajjad}
\email{redwan@virginia.edu}
\author{Avik W. Ghosh} 
\affiliation{Department of Electrical and Computer Engineering,
University of Virginia, VA 22904, USA.}

\date{\today}

\begin{abstract}
We explore the chiral transmission of electrons across graphene heterojunctions
for electronic switching using gate geometry alone. A sequence of gates
is used to collimate and orthogonalize the chiral transmission lobes across multiple junctions,
resulting in negligible overall current. The resistance of the device is enhanced by 
several orders of magnitude by biasing the gates into the bipolar $npn$ doping regime, as the ON state
in the near homogeneous $nn^-n$ regime remains highly conductive. The mobility is preserved because the 
switching involves a transmission gap instead of a structural band-gap 
that would reduce the number of available channels of conduction. Under certain conditions
this transmission gap is highly gate tunable, allowing a subthermal turn-on that beats the
Landauer bound on switching energy limiting present day digital electronics. 
\end{abstract}

\maketitle

The intriguing possibilities of graphene derive from its exceptional electronic and
material properties \cite{neto_09,sarma_10,beenakker_08}, in particular its photon-like bandstructure \cite{zhou_06}, ultrahigh mobility \cite{bolotin_08}, pseudospin physics and 
improved 2-D electrostatics
\cite{unluer_11}. Its switching ability, however, is compromised by the 
lack of a band-gap \cite{schwierz_10}, while opening a gap structurally kills the available
modes for conduction, degrading mobility \cite{tseng_10,schwierz_10}. This begs the question
as to whether we can significantly modulate the conductivity of graphene without any structural distortion, thereby preserving its superior mobility and electron-hole symmetry. 
A way to do this is to open a transmission gap that simply
redirects the electrons, without actually shutting off the density of states. The dual attributes
that help graphene electrons in this regard are its photon like trajectories and chiral tunneling that makes the junction resistance strongly anisotropic, allowing
redirection with gate geometry alone. 

In an earlier paper, \cite{sajjad_11} we outlined how we can open a transmission gap by a tunnel barrier, angularly injecting the electrons with a quantum point contact (QPC) and then selectively 
eliminating the low incidence angle Klein tunneling \cite{katsnelson_06} modes with a
barrier, in that case a patterned antidot or an insulating molecular chain. When the critical angle for total internal reflection is lower than the angle subtended at the QPC by the barrier, electrons are unable to cross over 
across the junction. The result is a transmission gap that can be collapsed by driving the voltage gradient across the junction towards the homogenous $pp$ or $nn$ limit, creating a subthermal turn-on sharper than the Landauer binary switching limit of $kTln2$ for distinguishability ($kTln10$ for each decade rise in current). Beyond proof of concept, that geometry was limited by a paucity of QPC modes and the structural distortions near the barrier that create a larger effective footprint. 

\begin{figure}[ht!]
\centering
\includegraphics[width=3.4in]{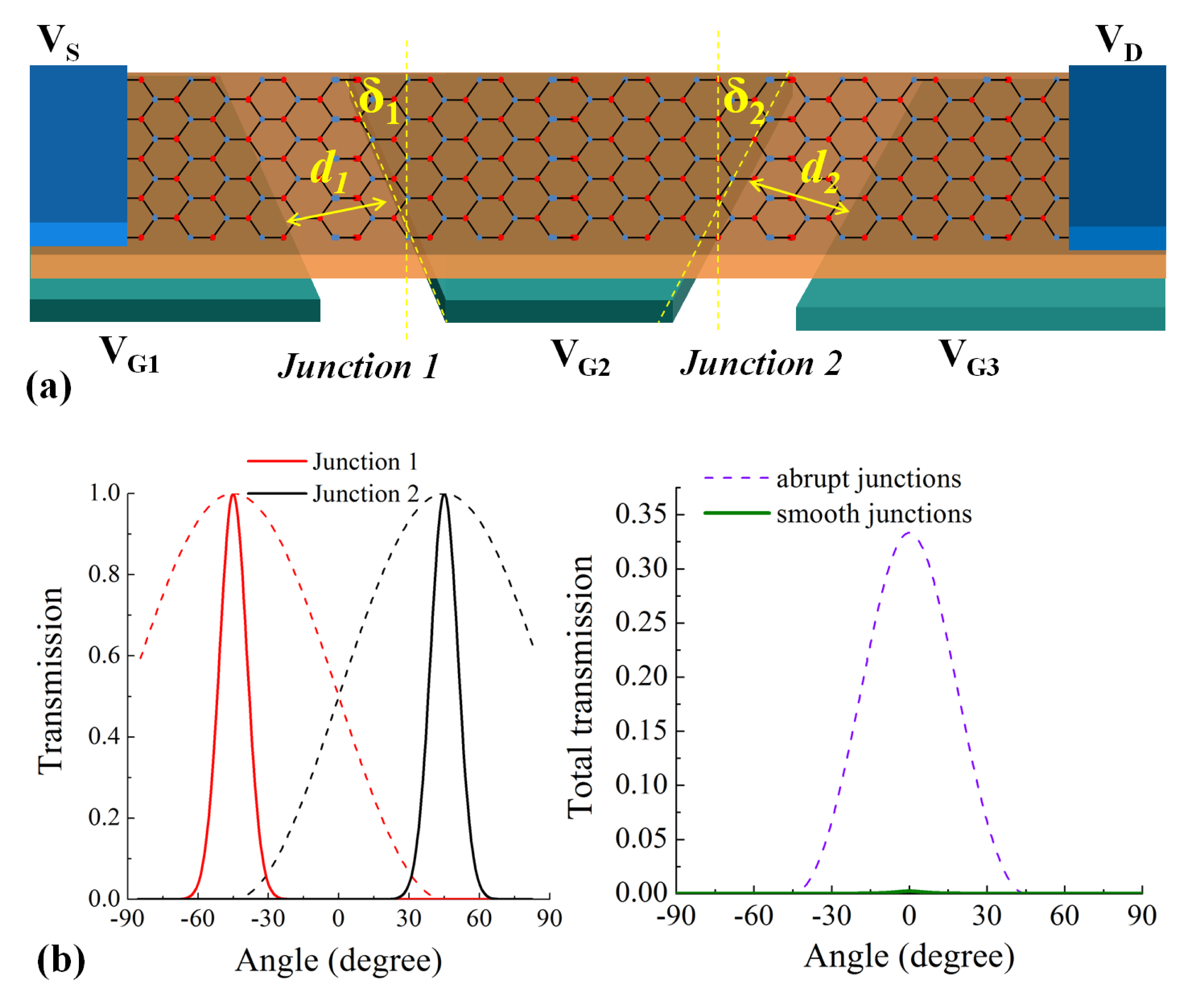}\quad
\subfigure{\includegraphics[width=2.9in]{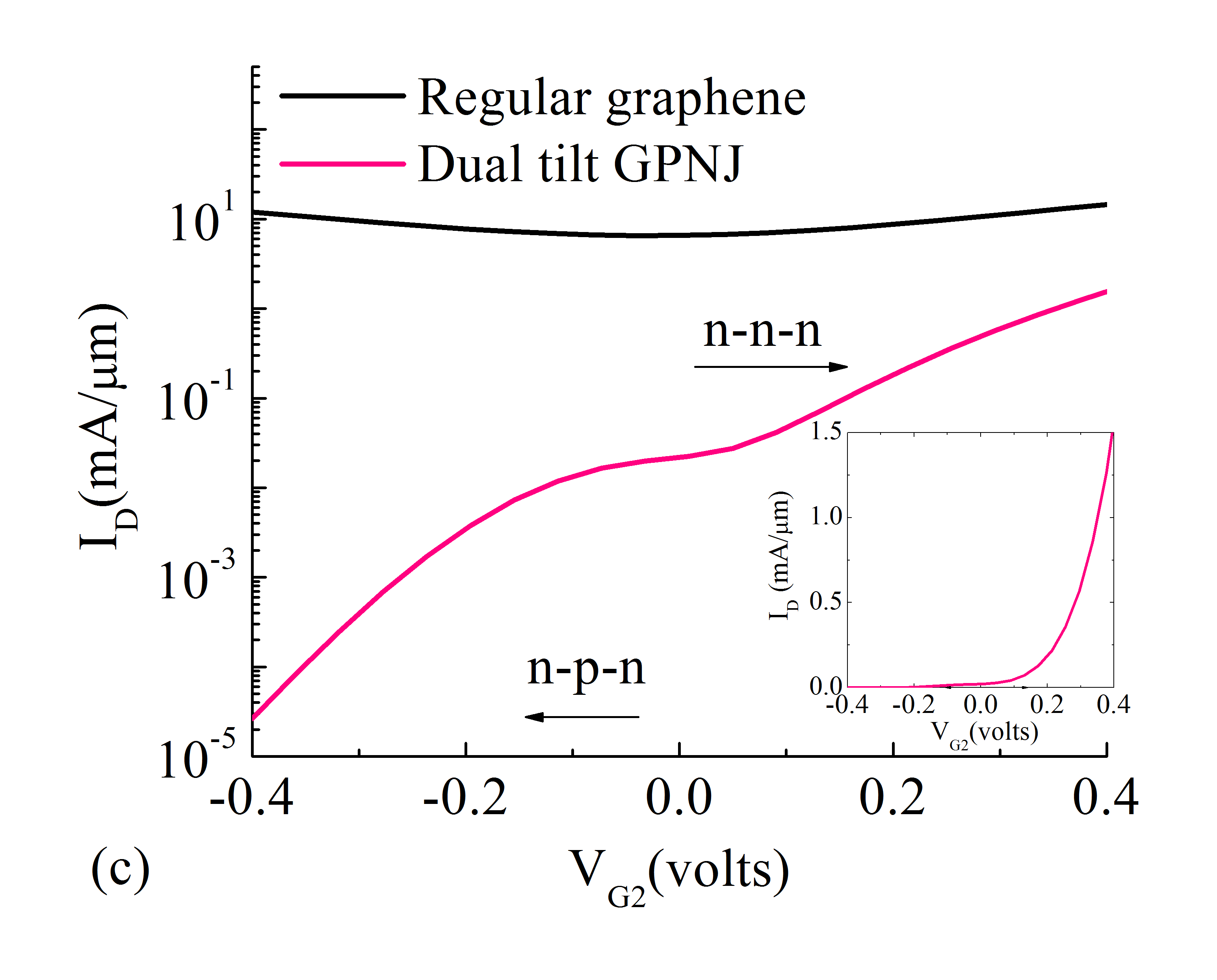}}\quad
\vskip -3mm
\caption{(Color online) (a) Chiral tunneling in graphene manipulated with gate geometry, using two junctions tilted in opposite directions, (b) making their angle dependent transmission lobes orthogonal (left) and yielding negligible overall transmission for well separated gates (right). (c) The transmission gap creates a high ON-OFF at finite bias, $V_{DS} = 0.4V$ and room temperature. The ON current degrades slightly compared to homogeneous gapless graphene, but the OFF current is reduced by several orders of magnitude.}
\label{nano0}
\end{figure}

In this paper, we combine a split gated $pn$ junction to collimate the transverse modes Fig. \ref{nano0}(a),
with recently demonstrated \cite{sutar_12, sajjad_12} action of a tilted $pn$ junction that increases the effective angle of incidence of the electrons. 
The conductance at zero temperature can be written as,
\begin{eqnarray}\label{eq1}
G(E_F) = G_0\sum_{n=1}^{M(E_F)} T_n = G_0\,\,M\,\,T_{av}
\end{eqnarray}where $G_0=2q^2/h$ is the conductance quantum for two spins, $M$ is the number of modes, $T_n$ is transmission of individual modes and $T_{av}$ is the average transmission over all modes. If all modes transmit with equal probability ($T$), the conductance can simply be written as $G_0MT$. Due to the chiral nature of carriers in graphene, transmission in GPNJ is highly angle (mode) dependent making it necessary to work with the average transmission per mode $T_{av}$. Instead of eliminating the mode count $M$ as does a structural band-gap, we 
exploit instead the chiral tunneling that makes $T_{av}$ vanishingly small over a range of energy and controllable with geometry alone (Fig.~\ref{nano0}b,c). All modes are available for conduction in the ON state when the split gates are set to the same polarity and thus retaining high mobility of graphene.

\textit{Engineering transmission gap with gate geometry.} Fig.~\ref{nano0} shows two $pn$ junctions tilted in opposite directions. Each junction exploits chiral tunneling that conserves pseudospin index and maximizes transmission at normal incidence (Klein tunneling), especially when they are smooth, i.e., the $p$ to $n$ transition occurs over a finite distance $2d$. A tilted junction rotates the transmission lobe accordingly, \cite{sajjad_12}, shifting transmissions along opposite directions to make them orthogonal. The mode-averaged transmissions across the dual junction can be decomposed as below (see appendix for details)
\begin{eqnarray}\label{dev_tr}
T_{1,2}(\theta) &\approx &  \Biggl[\frac{cos(\theta_L\pm \delta) cos\theta_R}{cos^2\biggl(\displaystyle\frac{\theta_L\pm \delta+\theta_R}{2}\biggr)}\Biggr]\nonumber \\
&\times& \exp{\biggl[\displaystyle -\pi d\frac{k_{FL}k_{FR}}{k_{FL} + k_{FR}} \sin{(\theta_L \pm \delta)}\sin{(\theta_R)}\biggr]}\nonumber\\
&&\nonumber\\
\displaystyle\frac{1}{T_{eff}} &\approx& \frac{1}{T_1} + \frac{1}{T_2} - 1 \\
T_{av}(E_F)&=& \frac{1}{2}\displaystyle \int T_{eff}(\theta)\cos{\theta}d\theta\nonumber\\
&=& [A\sqrt{k_Fd}e^{\pi k_Fdsin^2\delta}]^{-1}
\end{eqnarray} which is vanishingly small for moderate doping (Fermi wave-vector, $k_F = E_F/\hbar v_F$, $A$ is a constant, $A \approx 8$), gate split $2d$ and tilt angle $\delta$. The first equation arises from matching pseudospinors across the junction, $L$ and $R$ denoting components to left and right of a junction (1,2) . The tilt angle $\delta$ modifies the incident angle by $\theta_L\pm\delta$ and the angle of refraction is related to incident angle through Snell's law, $k_{FL}sin(\theta_L\pm\delta) = k_{FR}sin\theta_R$. The second equation assumes resistive addition of the junction resistances and ballistic flow in between. 
The mode count for an Ohmic contacted sample of width $W$ is given by $M = \frac{Wk_F}{\pi}$. The resulting total conductance $G_0MT_{av}$ is negligible in the entire $pn$ junction regime, indicating that the transmission gap ($E_G$) exists if the carrier densities have opposite polarities, 
\begin{eqnarray}\label{tgap}
E_G \approx V_0
\end{eqnarray} where $V_0$ is the gate induced voltage step across the junction. This is because the high resistance is primarily contributed by the WKB exponential factor which is valid in the $pn$ regime, whereas the unipolar regime has only the cosine prefactors represeting the wavefunction mismatch \cite{sajjad_13}. 

\begin{figure}
\centering
\includegraphics[width=3.4in]{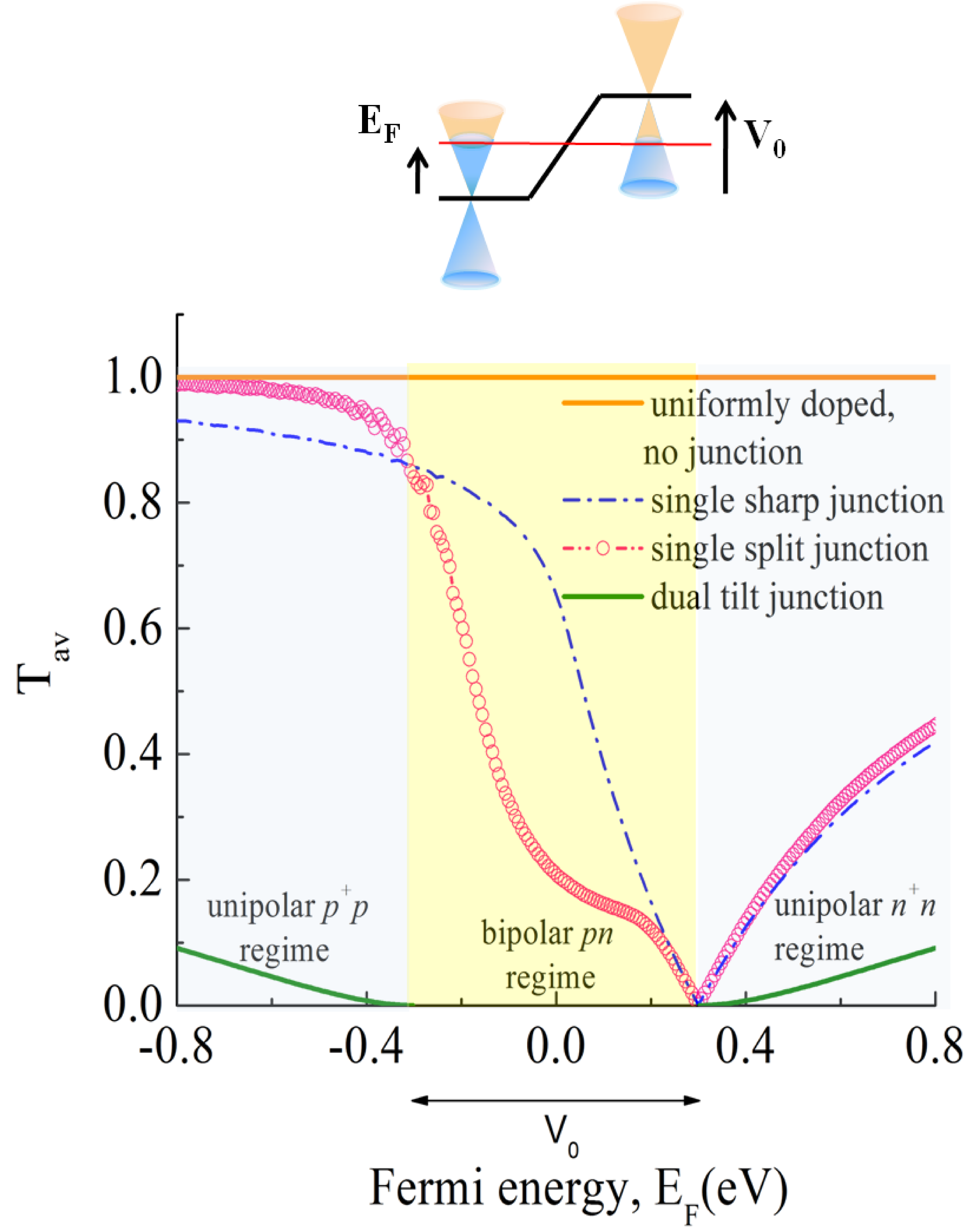}\quad
\caption{(Color online) Mode-averaged transmission $T_{av}$ vs Fermi energy $E_F$ for different doping profiles (Fermi energy $E_F$ and built in potential $V_0$ are indicated on the top band diagram). $T_{av}$ for the dual tilted GPNJ shows a gap (green line), which is termed as transmission gap (yellow shading) in this paper.}
\label{nano1}
\end{figure}
Fig. \ref{nano1} shows variation of $T_{av}$ numerically calculated from Eq.~\ref{dev_tr} as a function of Fermi energy ($E_F$) for four different devices and doping profiles. The orange line shows unit transmission of all modes for a ballistic uniformly doped graphene sheet. The angular (mode dependent) transmission is manifested in a single sharp ($d$ = 0) graphene $pn$ junction and the $T_{av}$ is suppressed (blue dots). Further suppression is achieved with a split junction (pink circles)  (non-zero $d$) due to high transverse energy (mode) filtering. $T_{av}$ for the device in Fig. \ref{nano1}(a) is shown in green, showing a negligible transmission over the bipolar doping regime. Note that both green and pink lines show suppression only in the bipolar doping regime, outside which the exponential scaling in Eq.\ref{dev_tr} is eliminated \cite{sajjad_13}. 

The minimum current is achieved in $npn$ regime (OFF state). Over the energy window $[\mu_S\,\,\,  \mu_D] = [E_F\,\,\, E_F-qV_{DS}]$ set by the drain voltage $V_{DS}$, $T_{av}$ varies weakly, so that the OFF state current at zero temperature for the $npn$ configuration can be extracted from
\begin{eqnarray}\label{current}
I_{OFF} = G_0\int_{\mu_D}^{\mu_S} M(E)T_{av}(E)dE\nonumber\\ \approx G_0M(E_F)T_{av}(E_F)V_{DS}
\end{eqnarray} convolved with the thermal broadening function at finite temperature. For uniformly doped graphene with ballistic transport,
\begin{eqnarray}
I_{ON} = G_0M(E_F)V_{DS}
\end{eqnarray} so that the zero temperature ON-OFF ratio simply becomes,
\begin{equation}\label{ONOFF}
I_{ON}/I_{OFF} \approx [T_{av}(E_F)]^{-1} \sim A\sqrt{k_Fd}(2e^{\displaystyle\pi k_Fdsin^2{\delta}})
\end{equation}
If the biasing is changed all the way from $npn$ to $nnn$. Fig. \ref{nano0}(c, pink line) shows the change in dual tilted GPNJ current with gate voltage $V_{G2}$ at room temperature and finite drain bias ($V_{DS}$), compared with a regular zero bandgap graphene based switch (black line). From the $nin$ to $nnn$ regime, we see little change in GPNJ current on a log scale. But towards the $npn$ regime, we see at least three orders of magnitude change when the Fermi window remains mostly within the transmission gap. Compared to the blue line, the ON current is reduced only slightly, while the OFF current is reduced by orders of magnitude. The reduction in ON current comes due to the fact that the doping is not quite uniform at the ON state across the $n^+n$ collimator (maintained at unequal doping to avoid a large voltage swing), whereupon the wave-function mismatch leads to lower current than usual. Fully ballistic transport assuming an Ohmic contacted high quality sample gives us an intrinsic ON current in the mA/$\mu\text{m}$ regime. In this calculation the gate parameters are $|\delta_1| = |\delta_2| = \delta = 45^0$, $d_1= d_2 = 20$nm, $V_{G1} = V_{G3} = +1$V, $V_{DS} = 0.4$V. 

\begin{figure}
\centering
\includegraphics[width=3.4in]{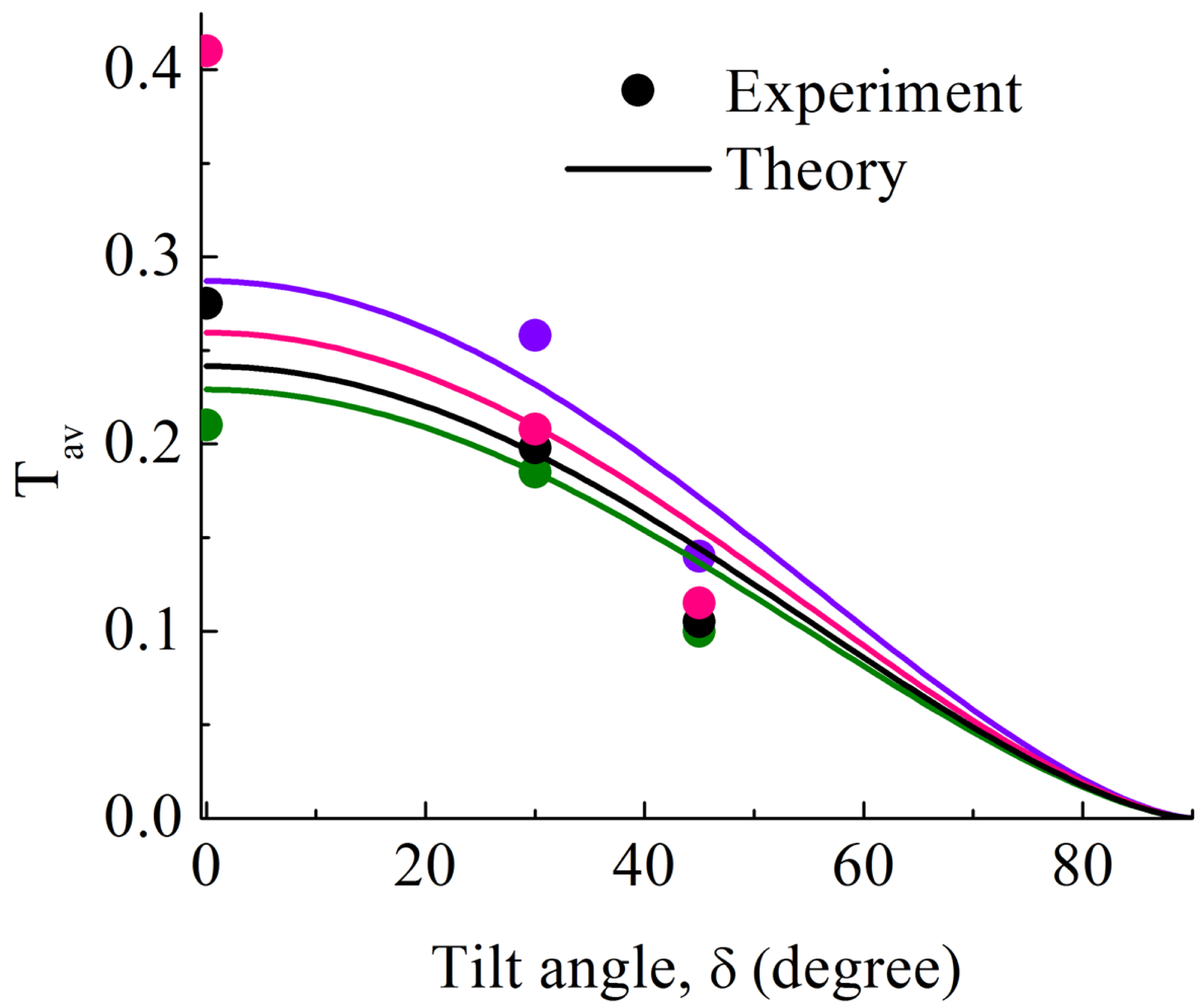}\quad
\caption{(Color online) Benchmarking $T_{av}$ with experiment \cite{sutar_12} for a single tilted split junction for several doping conditions. Experiment shows good agreement with the theory confirming the scaling law of tilt, Eq. \ref{tilt_scaling} .}
\label{nano_expt}
\end{figure}

Critical to the geometric switching is the prominance of angle-dependent chiral transmission across a tilted junction, especially in presence of charge puddles and edge reflection. Fig.~\ref{nano_expt} shows 
the mode averaged transmission extracted (see method in appendix) from the measured junction resistance for a single split junction, for varying tilt angles\cite{sutar_12}. For an abrupt tilted junction $T_{av} = 2/3 cos^4({\delta}/{2})$ in the symmetric $pn$ doping limit and represents an electronic analog of optical Malus' law. The reduction in $T_{av}$ happens due to the angular shift of transmission lobe (Fig. \ref{nano0}(b)) in low angular mode density region \cite{sajjad_12}. The numerically evaluated $T_{av}$ generalized for a tilted split junction (solid lines) agrees with experimental data (dots) from all the devices. This angular dependence persists, for multiple diffusive samples \cite{sutar_12}. 
The scaling of $T_{av}$ in experiment thus confirms the angular shift of the transmission lobes and forms the basis of the proposed device. The data show a remarkable absence of specular edge scattering, and can be explained by the randomizing effect of roughness.

\begin{figure*}
\centering
\includegraphics[width=6.8in]{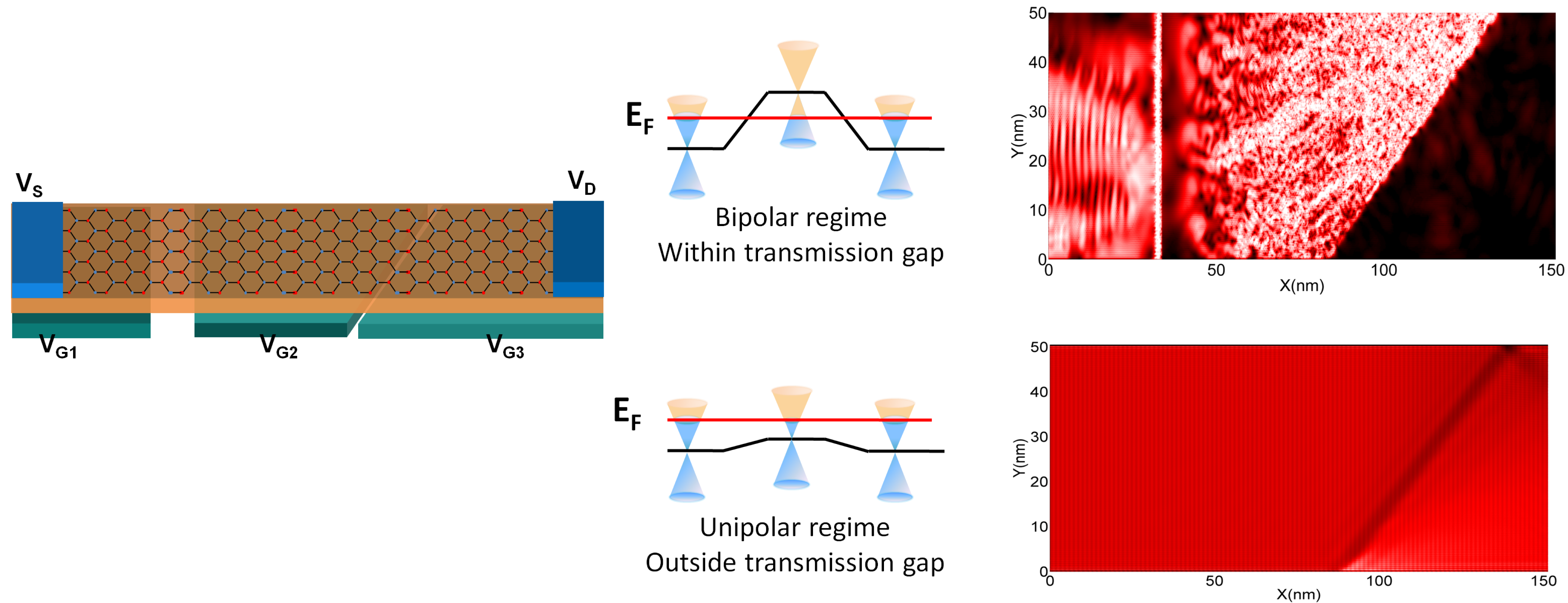}\quad
\caption{(Color online) Electron confinement in the proposed GPNJ device. Left: schematic of collimator-barrier pair that sequentially filters all propagating modes; middle column: band-diagrams showing bipolar OFF, ($npn$) vs unipolar ON, ($nn^{-}n$) states; (c) numerical current density plot from NEGF showing carrier reflections from the two junctions on top right (OFF), vs near uniform current flow at the bottom (ON). White (black) areas indicate high (low) local current density.}
\label{nano2}
\end{figure*}

\textit{Modified geometries and impact on subthreshold slope.} 
The aforementioned transmission gap is sensitive to gate parameters. In particular, making one of the junctions abrupt, using overlapping top and bottom 
gates, produces additional intricacies in addition to the high ON and low OFF current. 
Both the geometries in Figs.~\ref{nano0} and \ref{nano2} have $pn$ junctions aimed at
filtering out all propagating modes, but in Fig. \ref{nano2} the gate split $d_2 = 0$. The abruptness of the second junction makes the critical angle more sensitive to gate voltages and the transmission gap in Eq. \ref{tgap} needs to be changed. The first junction limits transmission primarily
to the Klein tunneling mode \cite{falko_06} in the OFF state, while the second junction, tilted at $\delta$ (Fig. \ref{nano2}(a)), increases the effective angle of incidence by the gate tilt angle $\delta$ \cite{sajjad_12}). 
All the electrons are then reflected if the critical angle of the second junction is less than $\delta$,
\begin{eqnarray}\label{critic}
\theta_C= \sin^{-1}|\frac{n_3}{n_2}| <\delta
\end{eqnarray}where $n_3$ and $n_2$ are doping concentrations on the two sides of junction 2. The resulting transmission vanishes over a range of energies (following from Eq.~\ref{critic}), which can be expressed as \cite{sajjad_11}
\begin{eqnarray}\label{trgap}
E_G = V_0\frac{{2sin\delta}}{{cos^2\delta}}
\end{eqnarray} analogous to Ref. \cite{sajjad_11} despite being a different (simpler) geometry, with the tilt angle $\delta$ replacing the barrier angle $\theta_B$.  

The tunability of the transmission gap for an abrupt junction bears a direct impact on the rate of change of current with gate voltage. For a semiconductor with fixed bandgap, this rate is 
$k_BTln(10)/q$ and limits the energy dissipation in binary switching. The limit arises from
the rate of change in overlap between the band-edge and the Fermi-Dirac distribution, normally
set by the Boltzmann tail. In our geometry however, the transmission gap is created artificially
with a gate bias $V_0$ across the GPNJ, and can be collapsed by going from heterogeneous ($npn$) towards the homogenous doping limit ($nnn$). Such a collapsible transport gap will overlap with the Fermi distribution at a higher rate than usual with change in gate bias, leading 
to a subthermal switching steeper than the Landauer limit. This results a lower gate voltage swing
to turn on the device and thus reducing dissipation.  


\textit{Numerical simulation of quantum flow.} To demonstrate carrier trajectories in the proposed device, we numerically solve the  
Non-Equilibrium Green's function Formalism (NEGF). The central quantity is the retarded Green's function,
\begin{eqnarray}\label{gr}
\mathcal{G} = (EI-H-U-\Sigma_1-\Sigma_2)^{-1}
\end{eqnarray} $H$ is the Hamiltonian matrix of graphene, described here with a minimal one $p_z$ orbital basis per carbon atom \textcolor{black}{with $t_0 = -3$eV being the hopping parameter.} $\Sigma_{1,2}$ are the self energy matrices for the semi-infinite source and drain leads, assumed to be extensions of the graphene sheet (i.e., assuming excellent contacts) and $\Gamma_{1,2}$ are the corresponding anti-Hermitian parts representing the energy level broadening  associated with charge injection and removal. $U$ is the device electrostatic potential. 
The current from $i$th atom to $jt$h atom is calculated from \cite{datta_97},
\begin{eqnarray}\label{crnt}
I_{i,j} = \frac{2q}{h}\int dE Im[\mathcal{G}^n_{i,j}(E)H_{j,i}-H_{i,j}\mathcal{G}^n_{j,i}(E)]
\end{eqnarray}where the electron correlation function, $\mathcal{G}^n=\mathcal{G}\Sigma^{in}\mathcal{G}^{\dagger}$ and in-scattering function, $\Sigma^{in} = \Gamma_Sf_S+\Gamma_Df_D$. The source and drain Fermi levels are at $\mu_S=0$ and $\mu_D = -qV_{DS}$. To see the current distribution in the device, we apply a small drain bias $V_{DS}$. $I_{i,j}$ is nonzero only if  the $i$th atom and $jt$h atom are neighbors. The total current at an atomic site can be found by adding all the components, $I_i = \sum_jI_{i,j}$. 

Fig. \ref{nano2} (right column) shows the local current density.
The $nn^{-}n$ ON state (bottom right) shows little reflection while the $npn$ OFF state (top right)
shows very little
current inside the final wedge connected to the drain. Most of the electrons that do not cross
the tilted junction are redirected towards the source by the edges. These electrons, especially
the secondary modes, are
rejected by the initial collimator and tend to build up in the central wedge. The
build-up of charge increases the local quasi-Fermi level $\mu$ until the injection rate at
the left junction, set by the transmission rate in Eq.~\ref{dev_tr}, equals the leakage rate at
the right tilted junction, given by the exponentially reduced tail in Eq.~\ref{ONOFF} plus
additional edge scattering based leakage pathways (a model was presented in \cite{sajjad_12} including a specularity
parameter $\eta$). 



In summary, by manipuilating the angle dependent chiral tunneling of GPNJ with patterned
gates alone, we can controllably suppress Klein tunneling and create a transmission gap as 
opposed to a structural band-gap. This is accomplished by combining the angular filtering at a split junction with the experimentally demonstrated chiral tunneling across tilted junctions, such that the transmission lobes across multiple junctions become orthogonal to each other in the OFF state. Since the ON
state simply requires changing the polarity of the central gate while sticking with otherwise pristine gapless graphene, the ON current stays very high. Furthermore, making the second junction abrupt renders its critical angle and thereby the overall transmission gap highly gate tunable, yielding a subthermal low-voltage turn-on that beats the Landauer switching limit. 
NEGF simulations show that the OFF current is limited by leakage aided by 
specular edge scattering, and is limited by build-up and blockade of charge in the central
angular wedge. 

\acknowledgements{The authors thank financial support from NRI-INDEX. The authors also thank Frank Tseng (UVa), Chenyun Pan (Georgia Tech), Azad Naeemi (Georgia Tech), Tony Low (IBM), Gianluca Fiori (U Pissa) for useful discussions.}
\bibliographystyle{apsrev4-1}

\begin{thebibliography}{16}%
\makeatletter
\providecommand \@ifxundefined [1]{%
 \@ifx{#1\undefined}
}%
\providecommand \@ifnum [1]{%
 \ifnum #1\expandafter \@firstoftwo
 \else \expandafter \@secondoftwo
 \fi
}%
\providecommand \@ifx [1]{%
 \ifx #1\expandafter \@firstoftwo
 \else \expandafter \@secondoftwo
 \fi
}%
\providecommand \natexlab [1]{#1}%
\providecommand \enquote  [1]{``#1''}%
\providecommand \bibnamefont  [1]{#1}%
\providecommand \bibfnamefont [1]{#1}%
\providecommand \citenamefont [1]{#1}%
\providecommand \href@noop [0]{\@secondoftwo}%
\providecommand \href [0]{\begingroup \@sanitize@url \@href}%
\providecommand \@href[1]{\@@startlink{#1}\@@href}%
\providecommand \@@href[1]{\endgroup#1\@@endlink}%
\providecommand \@sanitize@url [0]{\catcode `\\12\catcode `\$12\catcode
  `\&12\catcode `\#12\catcode `\^12\catcode `\_12\catcode `\%12\relax}%
\providecommand \@@startlink[1]{}%
\providecommand \@@endlink[0]{}%
\providecommand \url  [0]{\begingroup\@sanitize@url \@url }%
\providecommand \@url [1]{\endgroup\@href {#1}{\urlprefix }}%
\providecommand \urlprefix  [0]{URL }%
\providecommand \Eprint [0]{\href }%
\providecommand \doibase [0]{http://dx.doi.org/}%
\providecommand \selectlanguage [0]{\@gobble}%
\providecommand \bibinfo  [0]{\@secondoftwo}%
\providecommand \bibfield  [0]{\@secondoftwo}%
\providecommand \translation [1]{[#1]}%
\providecommand \BibitemOpen [0]{}%
\providecommand \bibitemStop [0]{}%
\providecommand \bibitemNoStop [0]{.\EOS\space}%
\providecommand \EOS [0]{\spacefactor3000\relax}%
\providecommand \BibitemShut  [1]{\csname bibitem#1\endcsname}%
\let\auto@bib@innerbib\@empty
\bibitem [{\citenamefont {Castro~Neto}\ \emph {et~al.}(2009)\citenamefont
  {Castro~Neto}, \citenamefont {Guinea}, \citenamefont {Peres}, \citenamefont
  {Novoselov},\ and\ \citenamefont {Geim}}]{neto_09}%
  \BibitemOpen
  \bibfield  {author} {\bibinfo {author} {\bibfnamefont {A.~H.}\ \bibnamefont
  {Castro~Neto}}, \bibinfo {author} {\bibfnamefont {F.}~\bibnamefont {Guinea}},
  \bibinfo {author} {\bibfnamefont {N.~M.~R.}\ \bibnamefont {Peres}}, \bibinfo
  {author} {\bibfnamefont {K.~S.}\ \bibnamefont {Novoselov}}, \ and\ \bibinfo
  {author} {\bibfnamefont {A.~K.}\ \bibnamefont {Geim}},\ }\href {\doibase
  10.1103/RevModPhys.81.109} {\bibfield  {journal} {\bibinfo  {journal} {Rev.
  Mod. Phys.}\ }\textbf {\bibinfo {volume} {81}},\ \bibinfo {pages} {109}
  (\bibinfo {year} {2009})}\BibitemShut {NoStop}%
\bibitem [{\citenamefont {Das~Sarma}\ \emph {et~al.}(2011)\citenamefont
  {Das~Sarma}, \citenamefont {Adam}, \citenamefont {Hwang},\ and\ \citenamefont
  {Rossi}}]{sarma_10}%
  \BibitemOpen
  \bibfield  {author} {\bibinfo {author} {\bibfnamefont {S.}~\bibnamefont
  {Das~Sarma}}, \bibinfo {author} {\bibfnamefont {S.}~\bibnamefont {Adam}},
  \bibinfo {author} {\bibfnamefont {E.~H.}\ \bibnamefont {Hwang}}, \ and\
  \bibinfo {author} {\bibfnamefont {E.}~\bibnamefont {Rossi}},\ }\href@noop {}
  {\bibfield  {journal} {\bibinfo  {journal} {Rev. Mod. Phys.}\ }\textbf
  {\bibinfo {volume} {83}},\ \bibinfo {pages} {407} (\bibinfo {year}
  {2011})}\BibitemShut {NoStop}%
\bibitem [{\citenamefont {Beenakker}(2008)}]{beenakker_08}%
  \BibitemOpen
  \bibfield  {author} {\bibinfo {author} {\bibfnamefont {C.~W.~J.}\
  \bibnamefont {Beenakker}},\ }\href@noop {} {\bibfield  {journal} {\bibinfo
  {journal} {Rev. Mod. Phys.}\ }\textbf {\bibinfo {volume} {80}},\ \bibinfo
  {pages} {1337} (\bibinfo {year} {2008})}\BibitemShut {NoStop}%
\bibitem [{\citenamefont {Zhou}\ \emph {et~al.}(2006)\citenamefont {Zhou},
  \citenamefont {Gweon}, \citenamefont {Graf}, \citenamefont {Fedrov},
  \citenamefont {Spataru}, \citenamefont {Diehl}, \citenamefont {Kopelevich},
  \citenamefont {Lee}, \citenamefont {Louie},\ and\ \citenamefont
  {Lanzara}}]{zhou_06}%
  \BibitemOpen
  \bibfield  {author} {\bibinfo {author} {\bibfnamefont {S.~Y.}\ \bibnamefont
  {Zhou}}, \bibinfo {author} {\bibfnamefont {G.-H.}\ \bibnamefont {Gweon}},
  \bibinfo {author} {\bibfnamefont {J.}~\bibnamefont {Graf}}, \bibinfo {author}
  {\bibfnamefont {A.~V.}\ \bibnamefont {Fedrov}}, \bibinfo {author}
  {\bibfnamefont {C.~D.}\ \bibnamefont {Spataru}}, \bibinfo {author}
  {\bibfnamefont {R.~D.}\ \bibnamefont {Diehl}}, \bibinfo {author}
  {\bibfnamefont {Y.}~\bibnamefont {Kopelevich}}, \bibinfo {author}
  {\bibfnamefont {D.~H.}\ \bibnamefont {Lee}}, \bibinfo {author} {\bibfnamefont
  {S.~G.}\ \bibnamefont {Louie}}, \ and\ \bibinfo {author} {\bibfnamefont
  {A.}~\bibnamefont {Lanzara}},\ }\href@noop {} {\bibfield  {journal} {\bibinfo
   {journal} {Nature}\ }\textbf {\bibinfo {volume} {2}},\ \bibinfo {pages}
  {595} (\bibinfo {year} {2006})}\BibitemShut {NoStop}%
\bibitem [{\citenamefont {Bolotin}\ \emph {et~al.}(2008)\citenamefont
  {Bolotin}, \citenamefont {Sikes}, \citenamefont {Jiang}, \citenamefont
  {Klima}, \citenamefont {Fudenberg}, \citenamefont {Hone}, \citenamefont
  {Kim},\ and\ \citenamefont {Stormer}}]{bolotin_08}%
  \BibitemOpen
  \bibfield  {author} {\bibinfo {author} {\bibfnamefont {K.}~\bibnamefont
  {Bolotin}}, \bibinfo {author} {\bibfnamefont {K.}~\bibnamefont {Sikes}},
  \bibinfo {author} {\bibfnamefont {Z.}~\bibnamefont {Jiang}}, \bibinfo
  {author} {\bibfnamefont {M.}~\bibnamefont {Klima}}, \bibinfo {author}
  {\bibfnamefont {G.}~\bibnamefont {Fudenberg}}, \bibinfo {author}
  {\bibfnamefont {J.}~\bibnamefont {Hone}}, \bibinfo {author} {\bibfnamefont
  {P.}~\bibnamefont {Kim}}, \ and\ \bibinfo {author} {\bibfnamefont
  {H.}~\bibnamefont {Stormer}},\ }\href@noop {} {\bibfield  {journal} {\bibinfo
   {journal} {Solid State Communications}\ }\textbf {\bibinfo {volume} {146}},\
  \bibinfo {pages} {351 } (\bibinfo {year} {2008})}\BibitemShut {NoStop}%
\bibitem [{\citenamefont {Unluer}\ \emph {et~al.}(2011)\citenamefont {Unluer},
  \citenamefont {Tseng}, \citenamefont {Ghosh},\ and\ \citenamefont
  {Stan}}]{unluer_11}%
  \BibitemOpen
  \bibfield  {author} {\bibinfo {author} {\bibfnamefont {D.}~\bibnamefont
  {Unluer}}, \bibinfo {author} {\bibfnamefont {F.}~\bibnamefont {Tseng}},
  \bibinfo {author} {\bibfnamefont {A.~W.}\ \bibnamefont {Ghosh}}, \ and\
  \bibinfo {author} {\bibfnamefont {M.~R.}\ \bibnamefont {Stan}},\ }\href@noop
  {} {\bibfield  {journal} {\bibinfo  {journal} {Nanotechnology, IEEE
  Transactions on}\ }\textbf {\bibinfo {volume} {10}},\ \bibinfo {pages} {931}
  (\bibinfo {year} {2011})}\BibitemShut {NoStop}%
\bibitem [{\citenamefont {Schwierz}(2010)}]{schwierz_10}%
  \BibitemOpen
  \bibfield  {author} {\bibinfo {author} {\bibfnamefont {F.}~\bibnamefont
  {Schwierz}},\ }\href@noop {} {\bibfield  {journal} {\bibinfo  {journal} {Nat
  Nano}\ }\textbf {\bibinfo {volume} {5}},\ \bibinfo {pages} {487} (\bibinfo
  {year} {2010})}\BibitemShut {NoStop}%
\bibitem [{\citenamefont {{Tseng}}\ and\ \citenamefont
  {{Ghosh}}(2010)}]{tseng_10}%
  \BibitemOpen
  \bibfield  {author} {\bibinfo {author} {\bibfnamefont {F.}~\bibnamefont
  {{Tseng}}}\ and\ \bibinfo {author} {\bibfnamefont {A.~W.}\ \bibnamefont
  {{Ghosh}}},\ }\href@noop {} {\bibfield  {journal} {\bibinfo  {journal} {ArXiv
  e-prints: 1003.4551}\ } (\bibinfo {year} {2010})},\ \Eprint
  {http://arxiv.org/abs/1003.4551} {arXiv:1003.4551 [cond-mat.mes-hall]}
  \BibitemShut {NoStop}%
\bibitem [{\citenamefont {Sajjad}\ and\ \citenamefont
  {Ghosh}(2011)}]{sajjad_11}%
  \BibitemOpen
  \bibfield  {author} {\bibinfo {author} {\bibfnamefont {R.~N.}\ \bibnamefont
  {Sajjad}}\ and\ \bibinfo {author} {\bibfnamefont {A.~W.}\ \bibnamefont
  {Ghosh}},\ }\href@noop {} {\bibfield  {journal} {\bibinfo  {journal} {Appl.
  Phys. Lett.}\ }\textbf {\bibinfo {volume} {99}},\ \bibinfo {pages} {123101}
  (\bibinfo {year} {2011})}\BibitemShut {NoStop}%
\bibitem [{\citenamefont {Katsnelson}\ \emph {et~al.}(2006)\citenamefont
  {Katsnelson}, \citenamefont {Novoselov},\ and\ \citenamefont
  {Geim}}]{katsnelson_06}%
  \BibitemOpen
  \bibfield  {author} {\bibinfo {author} {\bibfnamefont {M.~I.}\ \bibnamefont
  {Katsnelson}}, \bibinfo {author} {\bibfnamefont {K.~S.}\ \bibnamefont
  {Novoselov}}, \ and\ \bibinfo {author} {\bibfnamefont {A.~K.}\ \bibnamefont
  {Geim}},\ }\href@noop {} {\bibfield  {journal} {\bibinfo  {journal} {Nat
  Phys}\ }\textbf {\bibinfo {volume} {2}},\ \bibinfo {pages} {620} (\bibinfo
  {year} {2006})}\BibitemShut {NoStop}%
\bibitem [{\citenamefont {Sajjad}\ \emph {et~al.}(2012)\citenamefont {Sajjad},
  \citenamefont {Sutar}, \citenamefont {Lee},\ and\ \citenamefont
  {Ghosh}}]{sajjad_12}%
  \BibitemOpen
  \bibfield  {author} {\bibinfo {author} {\bibfnamefont {R.~N.}\ \bibnamefont
  {Sajjad}}, \bibinfo {author} {\bibfnamefont {S.}~\bibnamefont {Sutar}},
  \bibinfo {author} {\bibfnamefont {J.~U.}\ \bibnamefont {Lee}}, \ and\
  \bibinfo {author} {\bibfnamefont {A.}~\bibnamefont {Ghosh}},\ }\href@noop {}
  {\bibfield  {journal} {\bibinfo  {journal} {Phys. Rev. B}\ }\textbf {\bibinfo
  {volume} {86}},\ \bibinfo {pages} {155412} (\bibinfo {year}
  {2012})}\BibitemShut {NoStop}%
\bibitem [{\citenamefont {Low}\ and\ \citenamefont
  {Appenzeller}(2009)}]{low_tilted}%
  \BibitemOpen
  \bibfield  {author} {\bibinfo {author} {\bibfnamefont {T.}~\bibnamefont
  {Low}}\ and\ \bibinfo {author} {\bibfnamefont {J.}~\bibnamefont
  {Appenzeller}},\ }\href {\doibase 10.1103/PhysRevB.80.155406} {\bibfield
  {journal} {\bibinfo  {journal} {Phys. Rev. B}\ }\textbf {\bibinfo {volume}
  {80}},\ \bibinfo {pages} {155406} (\bibinfo {year} {2009})}\BibitemShut
  {NoStop}%
\bibitem [{\citenamefont {Sajjad}\ \emph {et~al.}(2013)\citenamefont {Sajjad},
  \citenamefont {Polanco},\ and\ \citenamefont {Ghosh}}]{sajjad_13}%
  \BibitemOpen
  \bibfield  {author} {\bibinfo {author} {\bibfnamefont {R.~N.}\ \bibnamefont
  {Sajjad}}, \bibinfo {author} {\bibfnamefont {C.}~\bibnamefont {Polanco}}, \
  and\ \bibinfo {author} {\bibfnamefont {A.~W.}\ \bibnamefont {Ghosh}},\
  }\href@noop {} {\bibfield  {journal} {\bibinfo  {journal} {arXiv preprint
  arXiv:1302.4473}\ } (\bibinfo {year} {2013})}\BibitemShut {NoStop}%
\bibitem [{\citenamefont {Sutar}\ \emph {et~al.}(2012)\citenamefont {Sutar},
  \citenamefont {Comfort}, \citenamefont {Liu}, \citenamefont {Taniguchi},
  \citenamefont {Watanabe},\ and\ \citenamefont {Lee}}]{sutar_12}%
  \BibitemOpen
  \bibfield  {author} {\bibinfo {author} {\bibfnamefont {S.}~\bibnamefont
  {Sutar}}, \bibinfo {author} {\bibfnamefont {E.~S.}\ \bibnamefont {Comfort}},
  \bibinfo {author} {\bibfnamefont {J.}~\bibnamefont {Liu}}, \bibinfo {author}
  {\bibfnamefont {T.}~\bibnamefont {Taniguchi}}, \bibinfo {author}
  {\bibfnamefont {K.}~\bibnamefont {Watanabe}}, \ and\ \bibinfo {author}
  {\bibfnamefont {J.~U.}\ \bibnamefont {Lee}},\ }\href@noop {} {\bibfield
  {journal} {\bibinfo  {journal} {Nano Letters}\ }\textbf {\bibinfo {volume}
  {12}},\ \bibinfo {pages} {4460} (\bibinfo {year} {2012})}\BibitemShut
  {NoStop}%
\bibitem [{\citenamefont {Cheianov}\ and\ \citenamefont
  {Fal'ko}(2006)}]{falko_06}%
  \BibitemOpen
  \bibfield  {author} {\bibinfo {author} {\bibfnamefont {V.~V.}\ \bibnamefont
  {Cheianov}}\ and\ \bibinfo {author} {\bibfnamefont {V.~I.}\ \bibnamefont
  {Fal'ko}},\ }\href@noop {} {\bibfield  {journal} {\bibinfo  {journal} {Phys.
  Rev. B}\ }\textbf {\bibinfo {volume} {74}},\ \bibinfo {pages} {041403}
  (\bibinfo {year} {2006})}\BibitemShut {NoStop}%
\bibitem [{\citenamefont {Datta}(1997)}]{datta_97}%
  \BibitemOpen
  \bibfield  {author} {\bibinfo {author} {\bibfnamefont {S.}~\bibnamefont
  {Datta}},\ }\href@noop {} {\emph {\bibinfo {title} {Electronic Transport in
  Mesoscopic Systems}}}\ (\bibinfo  {publisher} {Cambridge University Press},\
  \bibinfo {year} {1997})\BibitemShut {NoStop}%
\end{thebibliography}
%

\appendix*
\section{I}
\textit{The average tranmsission per mode:} Total transmission through a graphene heterojunction can be written as,
\begin{eqnarray}
G(E_F) &=& G_0\sum T(\theta)=G_0 \int \frac{T(\theta)}{\Delta \theta}d\theta\nonumber\\
&=&G_0 \frac{k_F}{\Delta k_y}\int T(\theta)cos\theta d\theta\nonumber\\
&=&G_0M(E_F)\frac{1}{2}\int T(\theta)cos\theta d\theta
\end{eqnarray}Here we have used, angular spacing, $\Delta \theta = \Delta k_y/(k_Fcos\theta)$, mode spacing $\Delta k_y = 2\pi/W$ and number of modes, $M(E_F) = Wk_F/\pi$. Comparing with Eq. \ref{eq1}, we can write,
\begin{eqnarray}\label{tr_av}
T_{av}(E_F) = \frac{1}{2}\int T(\theta)cos\theta d\theta
\end{eqnarray}Transmission through a single $pn$ junction, where the potential changes smoothely from $p$ to $n$ over a distance $2d$ is given by, 
\begin{eqnarray}
T(\theta) = e^{-\pi k_Fdsin^2\theta}
\end{eqnarray} ignoring the wave-function prefactor, this is valid for moderate gate split distance $2d$. 
Let us consider the $T_{av}$ for a single split junction and a tilted junction separately.

\begin{eqnarray}
G &\approx & G_0M(E_F)\frac{1}{2}\int_{-\theta_0}^{\theta_0}d\theta e^{-\pi k_Fd\theta^2}\nonumber\\
&=& G_0[\frac{1}{2\sqrt{k_Fd}}]M
\end{eqnarray} $T_{av}\approx \frac{1}{2\sqrt{k_Fd}}$ with gate split. 
For an abrupt tilted junction,
\begin{eqnarray}\label{delta}
G &\approx& G_0\int_{-\pi/2}^{\pi/2-\delta} \frac{T(\theta+\delta)}{\Delta \theta}d\theta \label{res_t}\nonumber\\&=& G_0[\frac{2}{3}cos^4(\frac{\delta}{2})] M\label{tilt_scaling}
\end{eqnarray} due to reduced density of modes at the higher angular region, $T_{av} =\frac{2}{3}cos^4(\frac{\delta}{2})$ is scaled with $\delta$. Therefore, a resistance measurement ($R_{Total} = 1/G$) will show an increase for a tilted device.\\

\textit{Transmission through dual tilt GPNJ device:}
In Fig. \ref{nano1}, we have two such junctions, each of them are tilted. Individual transmissions through the junctions becomes,
\begin{eqnarray}
T_1(\theta) = e^{-\pi k_Fdsin^2(\theta+\delta_1)}\\
T_2(\theta) = e^{-\pi k_Fdsin^2(\theta-\delta_2)}
\end{eqnarray}Since the tilt angle $\delta$ only modifies the angles of the incoming modes.

To get the total transmission, we combine the above two equations ignoring phase coherence to get the total transmission \cite{datta_97},
\begin{eqnarray}
\frac{1-T}{T} &=& \frac{1}{T_1}+\frac{1}{T_2}-2\nonumber\\
                    &=& e^{\pi k_Fdsin^2(\theta+\delta_1)}+e^{\pi k_Fdsin^2(\theta-\delta_2)}-2
\end{eqnarray}Overall transmission becomes
\begin{eqnarray}
T(\theta) = \frac{1}{e^{\pi k_Fdsin^2(\theta+\delta_1)}+e^{\pi k_Fdsin^2(\theta-\delta_2)}-1}
\end{eqnarray}And
\begin{eqnarray}
T_{av}(E_F)&=& \frac{1}{2}\int_{-\pi/2}^{\pi/2}\frac{d\theta cos\theta}{e^{\pi k_Fdsin^2(\theta+\delta_1)}+e^{\pi k_Fdsin^2(\theta-\delta_2)}-1}\nonumber\\
T_{av}(E_F) &\approx& \frac{1}{8}\frac{1}{\sqrt{k_Fd}(e^{\pi k_Fdsin^2\delta})}
\end{eqnarray}For $\delta_1 = \delta_2$. 

\textit{Extracting $T_{av}$ from transport measurement:} In the experiment \cite{sutar_12}, the junction resistance is extracted from
\begin{eqnarray}
Rj_{expt} = [R(V_{G1},V_{G2})+R(V_{G2},V_{G1}) \nonumber\\
-R(V_{G1},V_{G1})-R(V_{G2},V_{G2}]/{2},
\end{eqnarray} The above equation eliminates contact and device resistance due to scatterings and leaves out 
the resistance contribution from the $pn$ junction only. Theoretically the total resistance $R_{Total} = 1/G$ can be divided into two parts (contact and device resistance). From Eq. 1,
\begin{eqnarray}
R_{Total} &=& [G_0]^{-1}\frac{1}{MT_{av}}\\
 &=& [G_0]^{-1}[\frac{1}{M}+\frac{1-T_{av}}{MT_{av}}]
\end{eqnarray}In presence of a $pn$ junction with non-unity $T_{av}$, the second term can be considered as the junction resistance, 
\begin{eqnarray}\label{resist_junc}
Rj = [G_0]^{-1}[\frac{1-T_{av}}{MT_{av}}]
\end{eqnarray} While the theoretical $T_{av}$ is already known (Eq. \ref{res_t}), the experimental $T_{av}$ can be found by plugging the value of $Rj_{expt}$ from measurement in Eq. \ref{resist_junc} . The only unknown value remains is the number of modes at a particular gate voltage.
\begin{eqnarray}
M = \frac{W}{\pi}\frac{\Delta E(V_G)}{\hbar v_F}
\end{eqnarray}Here $\Delta E = \hbar v_F\sqrt{\pi C_GV_G/q}$ is the shift of Dirac point with gate voltage $V_G$. The gate capacitance is calculated from a simple parallel plate capacitor model $C_G = \frac{\epsilon}{t_{ox}}$where gate oxide thickness $t_{ox}$ is 100nm.

\end{document}